\documentclass[twocolumn,superscriptaddress,showpacs,prl,floatfix]{revtex4}
\usepackage{graphicx,epsfig,psfrag}
\usepackage{amsmath}
 
 \usepackage{wasysym} 
 \usepackage{mathrsfs} 
 \usepackage{graphicx} 
 \usepackage{amsfonts} 
 \usepackage{amsbsy} 
 \usepackage{amscd} 
 \usepackage{amssymb}
 \usepackage{pstricks} 
 \usepackage{multirow}

\newcommand{\be}{\begin{equation}}
\newcommand{\ee}{\end{equation}}
\newcommand{\bea}{\begin{eqnarray}}
\newcommand{\eea}{\end{eqnarray}}

\begin{document}
\title{
The lowest limits on the doubly charged Higgs boson masses \\in the minimal left-right symmetric
model 
}
\author{G. Bambhaniya}
\affiliation{Theoretical Physics Division, 
Physical Research Laboratory, \\
Navarangpura, Ahmedabad - 380009, India}
\author{J. Chakrabortty}
 \affiliation{Department of Physics, Indian Institute of Technology, Kanpur-208016, India}
 \author{  J. Gluza}
\author{T. Jeli\'nski}
\affiliation{Institute of Physics, University of Silesia, 
            Uniwersytecka 4, PL-40-007 Katowice, Poland }
 \author{M. Kordiaczy\'nska}
\affiliation{Institute of Physics, University of Silesia, 
            Uniwersytecka 4, PL-40-007 Katowice, Poland }
 
\begin{abstract} 
The doubly charged Higgs bosons $H_{1,2}^{\pm\pm}$ 
would undoubtedly be clear messengers 
of the new physics. We discuss their mass spectrum and show how experimental data and relations between scalar masses 
put limits on it. In particular, both the masses of the particles $H_1^0$, $A_1^0$ that  play a crucial role in FCNC effects and the masses of the additional gauge bosons  $W_2$, $Z_2$ 
are notably important. For instance, if $M_{H_1^0,A_1^0} \simeq 15$ TeV and $M_{W_2} \simeq 3.76$ TeV then the lowest mass of ${H_1^{\pm \pm}}$ is 465 GeV.  
In contrast, due to the freedom in the parameter space of the full scalar potential, there is no lowest limit on the mass of ${H_2^{\pm \pm}}$.  
It is shown to which signals at hadron colliders such  relatively light doubly charged scalars might correspond. 
LHC working at $\sqrt{s}=14$ TeV will enter into the region
where existence of such particles with minimal masses 
can be thoroughly explored for a much wider parameter 
space of the minimal and manifest version of the left-right symmetric model (MLRSM). Taking into account our considerations and present 
ATLAS and CMS exclusion limits on $M_{H^{\pm\pm}}$, there exist already first partial bounds on some of the MLRSM scalar potential parameters.

 \end{abstract}

\maketitle 
\allowdisplaybreaks

As argued among others in  \cite{Ellis:2013jnq},  the Higgs boson discovery at the LHC \cite{Aad:2012tfa,Chatrchyan:2012ufa} was a ``Big Deal". It established the Standard Model (SM) of gauge interactions and proved that the intuition  was correct, as far as the mechanism of mass generation is concerned. 
 
Now  further investigations are needed as we would like to know the details of the scalar sector. The underlying gauge theory need not to be necessarily the SM which is  based on the $SU(2)\times U(1)$ symmetry. In fact, there are many arguments that the SM is an effective theory working at relatively low energies \cite{Quigg:2009vq}. Then, going beyond the SM, one or two open and crucial questions are: what is the representation for the scalar multiplets and what is the shape of the scalar potential of the fundamental theory in particle physics? Are the fundamental forces unified, and if yes, then in which way? So far we do not have any particular answer to these queries. For instance, there are suggestions that the measured values of $m_H$ and $m_t$ make the electroweak vacuum metastable, and a new physics should be below $\sim10^{10}-10^{12}$ GeV \cite{Buttazzo:2013uya,Branchina:2013jra, Branchina:2014usa, Branchina:2014rva}. However, there are also arguments for the Standard Model being a valid theory up to the Planck scale \cite{Jegerlehner:2013cta}.   
 
SM scalar sector consists of a doublet of complex scalar field. There are four real scalar fields, three of them are responsible for giving masses to the three gauge bosons $W^\pm,Z^0$ through the Higgs mechanism. The remaining scalar is associated with the neutral Higgs boson which has been discovered at the LHC.
However, if we go beyond the SM, we may need to deal with more complex scalar systems, for instance $SU(2)$ triplet 
which contains charged (singly and(or) doubly) scalar particles.  These particles are naturally embedded in the left-right symmetric models where a new characteristic energy scale exists  and the symmetry between left and right $SU(2)$ 
gauge sectors is broken spontaneously \cite{Mohapatra:1974gc,Senjanovic:1975rk}. Here we focus on the so-called minimal and manifest version of the model (MLRSM), see e.g. \cite{Mohapatra:1974gc,Senjanovic:1975rk,Duka:1999uc,Mohapatra:1986uf}. 
%

There is presently large activity in searching for the scalar particles of any sort (neutral, charged) at the LHC. In particular, knowledge of the existence of doubly charged scalar bosons would be crucial for further directions in exploration of particle physics phenomena, for instance, their presence would strongly disfavour the minimal version of the supersymmetric Standard Model. There are two basic ways in which doubly charged particles can be searched for. The first is indirect: looking for rare lepton flavour and number violating processes
and precision measurements (deviations from the SM expectations). The second opportunity is provided by the accelerators where new particles can be produced directly at high energies, as in the LHC.   There are already many analysis undertaken by the CMS and ATLAS collaborations regarding these kind of searches,  and the present limits for doubly charged scalars are:
\begin{eqnarray}
M_{H^{\pm \pm}}\geq 445\; {\rm GeV} \,(409\; {\rm GeV})\;\;\; {\rm for} \;\;\; {\mbox{\rm  CMS (ATLAS)}},
\label{mhpplimit}
\end{eqnarray}
in the 100\% leptonic branching fraction scenarios \cite{CMS:2012kua,ATLAS:2012hi}. We assume the same framework here. Other  scenarios are also possible within MLRSM \cite{Bambhaniya:2013wza} or in another models, e.g. Higgs Triplet Model \cite{Melfo:2011nx,kang:2014jia}. 


In this paper, we show that theoretical considerations are also important, and we find that in the MLRSM some of the scalar masses are constrained from below. In particular, the lowest limit for the doubly charged scalar 
$H_1^{\pm \pm}$ mass is not much beyond the present limit given in Eq.~\ref{mhpplimit} for many different model parameters.  

Recently,  it has been shown in \cite{Bambhaniya:2013wza} that charged scalars with masses at the level of a few hundred GeV
can be realized in the MLRSM  by allowing a percent-level tuning of the scalar potential parameters, compatible with both the large parity breaking scale $v_R$ and severe bounds on neutral scalar masses ($M_{H_1^0,A_1^0}$) derived from flavour changing neutral currents (FCNC). This is non-trivial as both $v_R$ and $M_{H_1^0,A_1^0}$ are at the level of a few $\mathrm{TeV}$ and all scalars apart form $H_0^0$ are naturally heavy, their leading mass terms are proportional to 
$v_R$.  
In this paper, we look into further details and show that in fact lowest bounds can be obtained for some of the scalar masses. The MLRSM scalar potential and its minimization followed by the diagonalization have been investigated in  \cite{Gunion:1989in} and explicit relations between physical and unphysical scalar fields are given in \cite{Gluza:1994ad}. 
For our purposes, we repeat here only a subset of formulas which we need for further discussion, they are valid 
as long as $\kappa_1\ll v_R$, which is true as $\kappa_1$ and $v_R$ are connected  directly with masses of light and heavy charged gauge bosons, and $M_{W_1}\ll M_{W_2}$ \cite{Beringer:1900zz,Bambhaniya:2013wza} 
\begin{eqnarray}
M^2_{H^0_0} &\simeq & 2\kappa^2_1
\lambda_1,\\ 
M^2_{H^0_1}&\simeq & \frac{1}{2} \alpha_3 v^2_R, \label{mass1}\\
M^2_{A^0_1} &\simeq& \frac{1}{2} \alpha_3 v^2_R
-2\kappa^2_1 \left(
2\lambda_2-\lambda_3 \right), \label{ma10}\\
M^2_{H^0_3} &\simeq & \frac{1}{2} v^2_R
\left(
\rho_3 - 2\rho_1
\right), \label{mh30}\\
M^2_{H_1^{\pm \pm}} &\simeq& \frac{1}{2}
\left[v^2_R
\left(\rho_3-2\rho_1
\right)+\alpha_3\kappa^2_1\right],\label{hpp1}\\
M^2_{H_2^{\pm \pm}} &\simeq & 2\rho_2 v^2_R+
\frac{1}{2} \alpha_3 \kappa^2_1. 
\label{hpp2} 
\end{eqnarray} 

As we can see, SM-like Higgs boson $H_0^0$ has a  mass proportional to the vacuum expectation value $\kappa_1$ ($\sim$ electroweak breaking scale,
$M_{W_1} \sim \kappa_1$ \cite{Duka:1999uc}). Here, among the neutral scalars only $A_1^0$ and $H_1^0$ contribute to the FCNC interactions. To our knowledge, their effects have been discussed for the first time in the context of left-right models in \cite{Ecker:1983uh}, see also \cite{Mohapatra:1983ae,Pospelov:1996fq,Zhang:2007da,Maiezza:2010ic,Chakrabortty:2012pp,Bertolini:2014sua} and the Appendix in \cite{Bambhaniya:2013wza}. In general their masses need to be at least of the order of 10 TeV, though some alternatives have been also considered in \cite{Guadagnoli:2010sd}. Also, the parity breaking scale $v_R$ is already strongly constrained by a 
``golden" decay chain process $W_R \to l_1 N_l \to l_1 l_2 jj$  \cite{CMS:2012uaa,salo:1558322}, 
$M_{W_2} \geq  2.8$ {\rm TeV}. This limit (at 95 \% C.L.) is for a genuine left-right symmetric model which we consider here (MLRSM). 
There are also other model independent limits on the masses of gauge bosons $\sim$ 4 TeV and we stay conservative while choosing this value that implies $v_R$ is already bigger than 8 TeV. For more details and other possible scenarios
see \cite{Bambhaniya:2013wza} and recent \cite{Deppisch:2014qpa,Heikinheimo:2014tba}.
 
Now, taking into account the above facts, for example with $M_{H_1^0,A_1^0} \geq 10$ TeV and $v_R \geq 8$ TeV, the scalar potential parameter $\alpha_3$ can be determined, see Eq.~\ref{mass1}. This parameter also enters into the relations of the masses of doubly charged scalars, Eqs.~\ref{hpp1} and \ref{hpp2}. Thus the only free parameter that constrains $M_{H_1^{\pm \pm}}$ is $\delta \rho \equiv \rho_3-2 \rho_1$.

\begin{figure}[h!]
\epsfig{figure=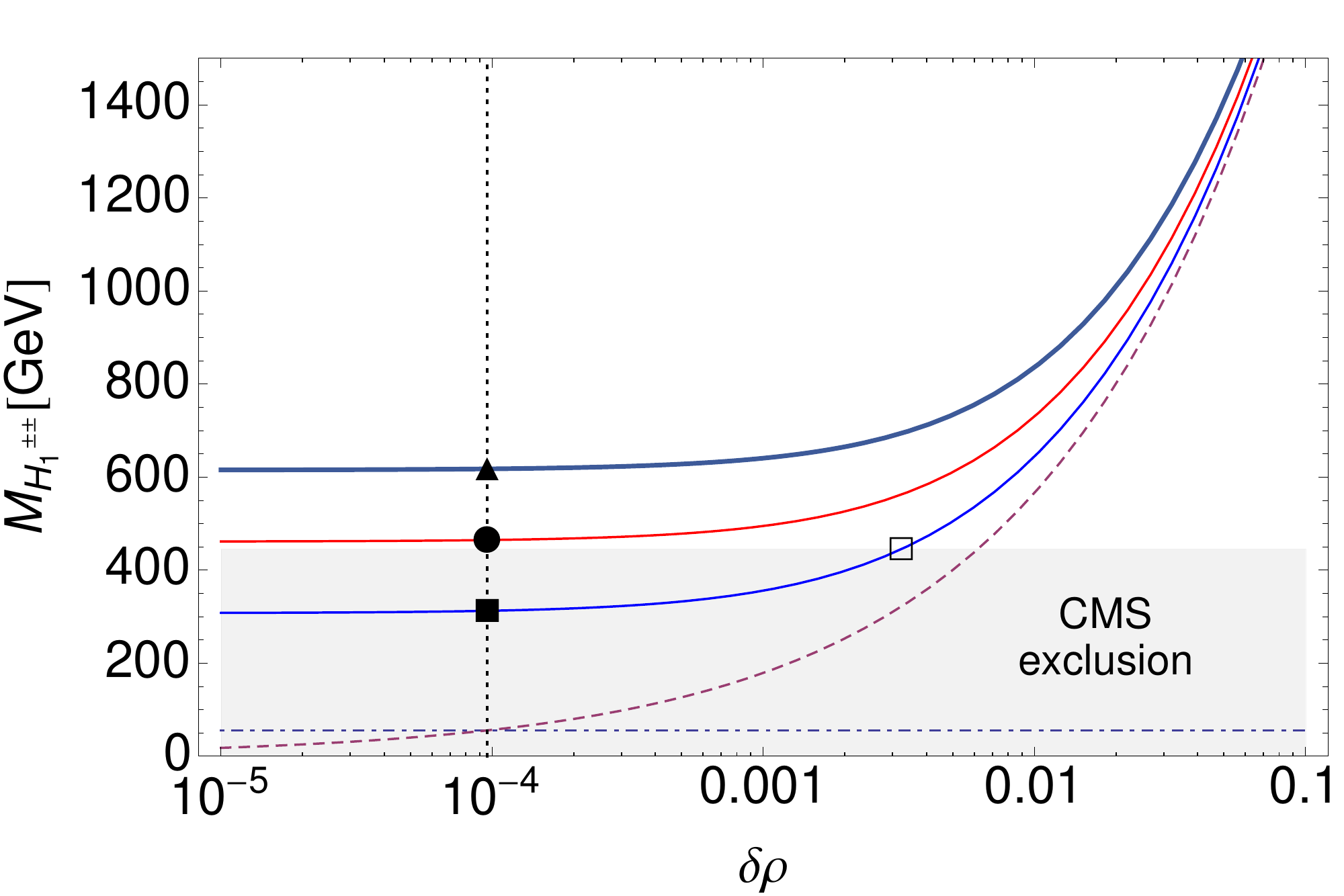,
height=5.75cm}
\caption{\label{fig1}  Dependence of the ${H_1^{\pm \pm}}$ mass on $\delta \rho \equiv \rho_3-2 \rho_1$ for $v_R=8\,\mathrm{TeV}$ and various masses of flavour changing neutral Higgs scalars 
$H_1^0$ and $A_1^0$. Points marked on the vertical dotted line by square, circle and triangle show minimal $M_{H_1^{\pm\pm}}$ corresponding to $M_{H_1^0,A_1^0}=10$, $15$ and $20$ TeV, respectively (see Tab.~\ref{table1}). In the region left to the vertical dotted line, $M_{H_3^0}$ (dashed line) is smaller than $55.4\,\mathrm{GeV}$ (dash-dotted horizontal line). The horizontal grey strip represents the latest CMS exclusion limit on doubly charged Higgs boson mass. The mass spectrum represented by the filled square is obviously ruled out by the LHC. Hence, for $v_R=8\,\mathrm{TeV}$ and $M_{H_1^0,A_1^0}=10\,\mathrm{TeV}$ the minimal allowed value of $M_{H_1^{\pm\pm}}$ is 445 GeV (empty square) and one gets the following lower bound $\delta\rho\gtrsim3\cdot10^{-3}$. 
}
\end{figure}

In Fig.~\ref{fig1} mass of $H_1^{\pm \pm}$ is given as a function of $\delta \rho$ for $v_R=8\,\mathrm{TeV}$ and different choices of $M_{H_1^0,A_1^0}$. Without the loss of generality it has been assumed that $M_{H_1^0}=M_{A_1^0}$, which implies that $2 \lambda_2=\lambda_3$, Eq.~\ref{ma10}.  
Here, $\delta \rho$ starts from zero, as we aim at the minimal values of $M_{H_1^{\pm \pm}}$.
For  $\delta \rho=0$, the only contribution to the 
$H_1^{\pm \pm}$ mass is connected with a second term in Eq.~\ref{hpp1}, however, also mass of $H_3^0$ 
depends directly on $\delta \rho$, see Eq.~\ref{mh30}. Thus it cannot be negative. Moreover, interpreting LEP II data  related to $e^+e^-\rightarrow\gamma+E_T\!\!\!\!\!\!\!{\slash}\,\,\,\,$, it is possible to find lower bound on $M_{H_3^0}$, which is about $55.4\,\mathrm{GeV}$ \cite{Datta:1999nc}. 
This leads to the minimal value for the mass of doubly charged scalar ${H_1^{\pm \pm}}$.  It depends both on the  minimal allowed values of $M_{H_3^0}$ and masses of the scalars ${H_1^0,A_1^0}$ that control FCNC. Let us note that for $v_R\sim10\,\mathrm{TeV}$ and  small values of
$\delta\rho$ (e.g. $\delta\rho\lesssim5\cdot10^{-4}$), the dependence on $\delta\rho$ is rather weak and the mass of $H_1^{\pm\pm}$ is dominated by the term 
$\sqrt{\alpha_3}\kappa_1/\sqrt{2}\approx171\sqrt{\alpha_3}\,\;\mathrm{GeV}$. 

As one can see from Fig.~\ref{fig1}, for $M_{H_1^0,A_1^0}=10\,\mathrm{TeV}$ bounds from the LHC yield the following limit: $\delta\rho\gtrsim3\cdot10^{-3}$. On the other hand, for bigger $M_{H_1^0,A_1^0}$ there are no restrictions on $\delta\rho$ from the LHC yet but the model itself predicts the minimal possible value of $M_{H_1^{\pm\pm}}$ masses. 

It is worthwhile to note that taking into account experimental limits one can obtain the following bound on $M_{H_1^{\pm\pm}}$:
\begin{equation}\label{MH1ppvR}
\sqrt{\min(M_{H_3^0}^2)+M_{H_1^0}^2\kappa_1^2/v_R^2}\approx\frac{2.41\;\mathrm{TeV}^2}{v_R}.
\end{equation}

The dashed line in Fig.~\ref{fig1} is for $M_{H_3^0}$. At $\delta\rho\approx10^{-4}$, it crosses the horizontal line which corresponds to the lowest bound on mass of $H_3^0$. Hence, the region left to the vertical, dotted line is excluded due to the above mentioned LEP constraints 
and 
minimal mass of  ${H_1^{\pm \pm}}$ can be determined from the points at which solid lines intersect vertical, dotted line. Some precise values of ${H_1^{\pm \pm}}$ masses corresponding to such points are presented in the Tab.~\ref{table1}. 

\begin{table}[h!]
\begin{tabular}{|c|c|c|}
\hline 
\hline  
\multicolumn{3}{|c|}{$v_R=8$ TeV ($M_{W_2}=3.76$ TeV)}\\
\hline 
$M_{A_1^0,H_1^0}$ [TeV]& $\alpha_3$& $\mathrm{min}(M_{H_1}^{\pm \pm})$ [GeV]\\
\hline
10  &  3.13 & 312 ({\tiny $\blacksquare$})\\
\hline
15  &  7.03 & 465 ({\tiny $\CIRCLE$})\\
\hline
20  &  12.5 & 617 ({\footnotesize $\blacktriangle$})\\
\hline
\multicolumn{3}{|c|}{$v_R=10$ TeV ($M_{W_2}=4.7$ TeV)}\\
\hline
$M_{A_1^0,H_1^0}$ [TeV]& $\alpha_3$& $\mathrm{min}(M_{H_1}^{\pm \pm})$ [GeV]\\
 \hline
10  &  2.0 & 252\\
\hline
15  &  4.5 & 373\\
\hline
20  &  8.0 & 495\\
\hline
\multicolumn{3}{|c|}{$v_R=12$ TeV ($M_{W_2}=5.64$ TeV)}\\
\hline
$M_{A_1^0,H_1^0}$ [TeV]& $\alpha_3$& $\mathrm{min}(M_{H_1}^{\pm \pm})$ [GeV]\\
\hline
10  &  1.39 & 212\\
\hline
15  &  3.13 & 312\\
\hline
20  &  5.56 & 413\\ 
\hline
\hline
\end{tabular}
\caption{Minimal masses of a doubly charged Higgs boson $H_1^{\pm\pm}$  as a function of the parity breaking scale   $v_R$ of the right sector of the model and the mass of neutral Higgs bosons $ \{A_1^0,H_1^0\}$ which contribute to the FCNC effects. 
Corresponding masses of $H_2^{\pm\pm}$ are fixed by taking in addition $\rho_2=\delta\rho/4$, what yields $M_{H_2^{\pm\pm}}=M_{H_1^{\pm\pm}}$. No LHC direct limits from Eq.~\ref{mhpplimit} applied. Square, circle and triangle correspond to points marked on Fig.~\ref{fig1}.}
\label{table1}
\end{table}

Let us remark that there is, in principle, no restriction from the scalar potential parameters on $M_{H_2^{\pm \pm}}$, as the parameter $\rho_2$ does not play any role to determine the masses of other remaining scalars, $H_2^0, A_2^0, H_1^\pm, H_2^\pm$. Thus this parameter is not constrained within perturbative limit. Here, for example we have set $\rho_2=\delta\rho/4$ in Tab.~\ref{table1}. Moreover, note that $\min(M_{H_{1}^{\pm\pm}})$ decreases with $v_R$, see Eq.~\ref{MH1ppvR}.

\begin{figure}[h!]
\epsfig{figure=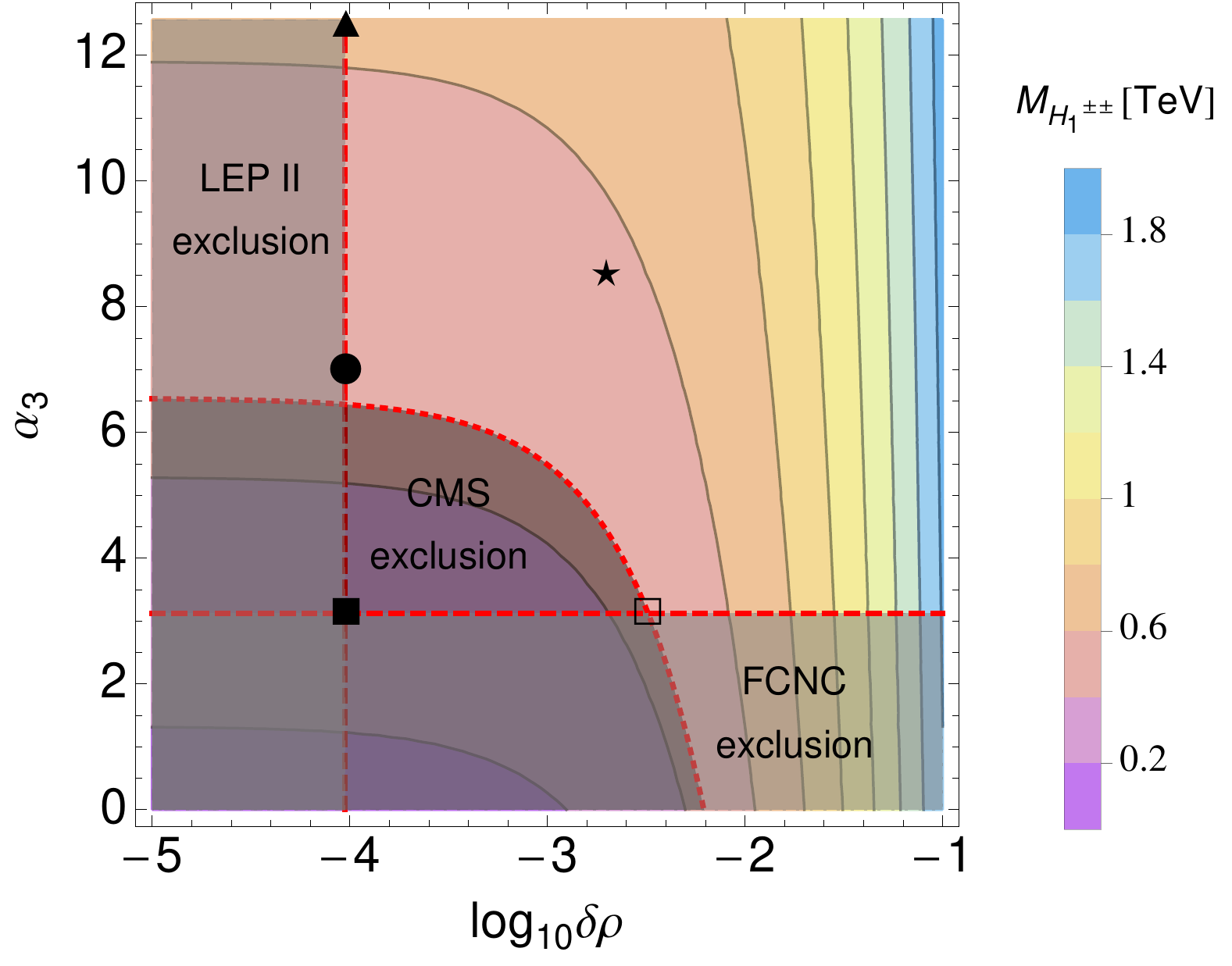,
scale=0.545}
\caption{\label{fig1b}  Dependence of the ${H_1^{\pm \pm}}$ mass (in TeV) on $\delta \rho$ and $\alpha_3$ for $v_R=8\,\mathrm{TeV}$. Solid lines divide the parameter space into colored regions in which mass of $H_1^{\pm\pm}$ is characterized according to the legend. Point marked by a $\star$ corresponds to benchmark set of parameters (\ref{BP2a})-(\ref{BP2b}). Shaded regions are excluded due to FCNC (under horizontal, dashed line), LEP (left to vertical dashed line) and CMS constraints (under curved dotted line). Squares, circle and traingle correspond to points marked in Fig.~\ref{fig1}.
}
\end{figure}

It is clear that some of $M_{H_1^{\pm\pm}}$ in Tab.~\ref{table1} are already excluded by the LHC limits, Eq.~\ref{mhpplimit}. 
It means that the first constraints on scalar potential parameters can be derived from those experimental bounds, as already seen in Fig.~\ref{fig1} (point marked by empty square). 
To this end, it is worthwhile to examine the influence of $\alpha_3$ on masses of  $H_1^{\pm\pm}$. 
Let us focus on $v_R=8\,\mathrm{TeV}$ case. When $\delta\rho$ is smaller than about $10^{-4}$ then mass of doubly charged scalar is dominated by the term $\sqrt{\alpha_3}\kappa_1/\sqrt{2}$. On the other hand, if $\delta\rho$ is bigger than $10^{-2}$ then its mass is mainly driven by $v_R\sqrt{\delta\rho/2}$. On Fig.~\ref{fig1b}, 
it is shown how those masses depend on $\delta\rho$ and $\alpha_3$ in the region of the parameter space where both contributions to masses are important. On that
plot, we have shaded the regions of the parameter space which are excluded either by FCNC (i.e. $M_{H_1^0,A_1^0}\gtrsim10\,\mathrm{TeV}$) or LHC. Note that 
there is also excluded region related to the lower bound on $H_3^0$ mass (left of the vertical dashed line).
 It is straightforward to do similar analysis for dependence of $H_2^{\pm\pm}$ mass on $\alpha_3$ and $\rho_2$.

As an example, we present benchmark set of scalar masses which satisfy both FCNC and LHC constraints with 
light and degenerate masses of doubly charged scalars, assuming $v_R=8$ TeV (all masses are given in GeV):
%
\begin{eqnarray}
M_{H^0_0} &=& 125,\label{B2MH00}\\ 
M_{H^0_1}&= & 16492,\;\;\;
M_{H^0_2} = 11314, \;\;\;
M_{H^0_3} = 253, \\
M_{A^0_1} &=& 16496,\;\;\;
M_{A^0_2} =  253,\\
M_{H^\pm_1} &=& 439,\;\;\;
M_{H^\pm_2} = 16496,  \\
M_{H_1^{\pm \pm}} &=& 567, \;\;\;
M_{H_2^{\pm \pm}} = 567.\label{B2MH2pp}
\end{eqnarray} 
These masses are outcome of the following parameters of the scalar potential:
\begin{eqnarray}\label{BP2a}
\rho_1 &=& 1.0,\;\;\;\rho_2=5\cdot10^{-4},\;\;\;\rho_3-2\rho_1=2\cdot10^{-3},\\
\lambda_1 &=& 0.13,\;\;\; \lambda_2=0,\;\;\;\lambda_3=1,\\
\alpha_3&=& 8.5.\label{BP2b}
\end{eqnarray}

This scenario realizes mass spectrum with maximal number (three) of light charged scalars \cite{Bambhaniya:2013wza}. Of course, yet another setups  are also possible.  
%
For example, increasing parity breaking scale to $20\,\mathrm{TeV}$, setting $\alpha_3\sim0.5$, $\rho_1\sim\rho_2\gtrsim\alpha_3/4$ and tuning $\delta\rho\sim10^{-3}$ give $H_3^0$, $A_2^0$, $H_1^{\pm}$ and $H_1^{\pm\pm}$ masses $\sim v_R\sqrt{\delta\rho/2}\sim450\,\mathrm{GeV}$, while the remaining scalars (beside $H_0^0$) have masses larger than $v_R\sqrt{\alpha_3/2}\sim10\,\mathrm{TeV}$. 
The other option is to keep $v_R\sim10\,\mathrm{TeV}$ but also choose $\rho_2\sim10^{-3}$ and set $\delta\rho\sim\alpha_3\sim2$. As a result, only $H_2^{\pm\pm}$ is light, with mass $\sim500\,\mathrm{GeV}$, while other scalars are heavier than $10\,\mathrm{TeV}$.

Let us note that all these spectra are also in agreement with low-energy constraints, here muon $\Delta r$ parameter plays a crucial role \cite{Czakon:1999ue,Czakon:2002wm,Chakrabortty:2012pp}. It is interesting to note that scalar and fermionic loop contributions come up  with opposite signs to $\Delta r$ and heavy neutrinos can compensate possible large scalar effects \cite{Czakon:1999ue,Czakon:2002wm,Chakrabortty:2012pp}. 

We have focused here on doubly charged Higgs bosons, aiming at their small masses, but we can see that they are entangled by the  scalar potential parameters, and in this way also some singly charged and neutral scalars are relatively light. Their effects at hadron colliders will be studied in details elsewhere. 

\begin{figure}[htb]
\epsfig{figure=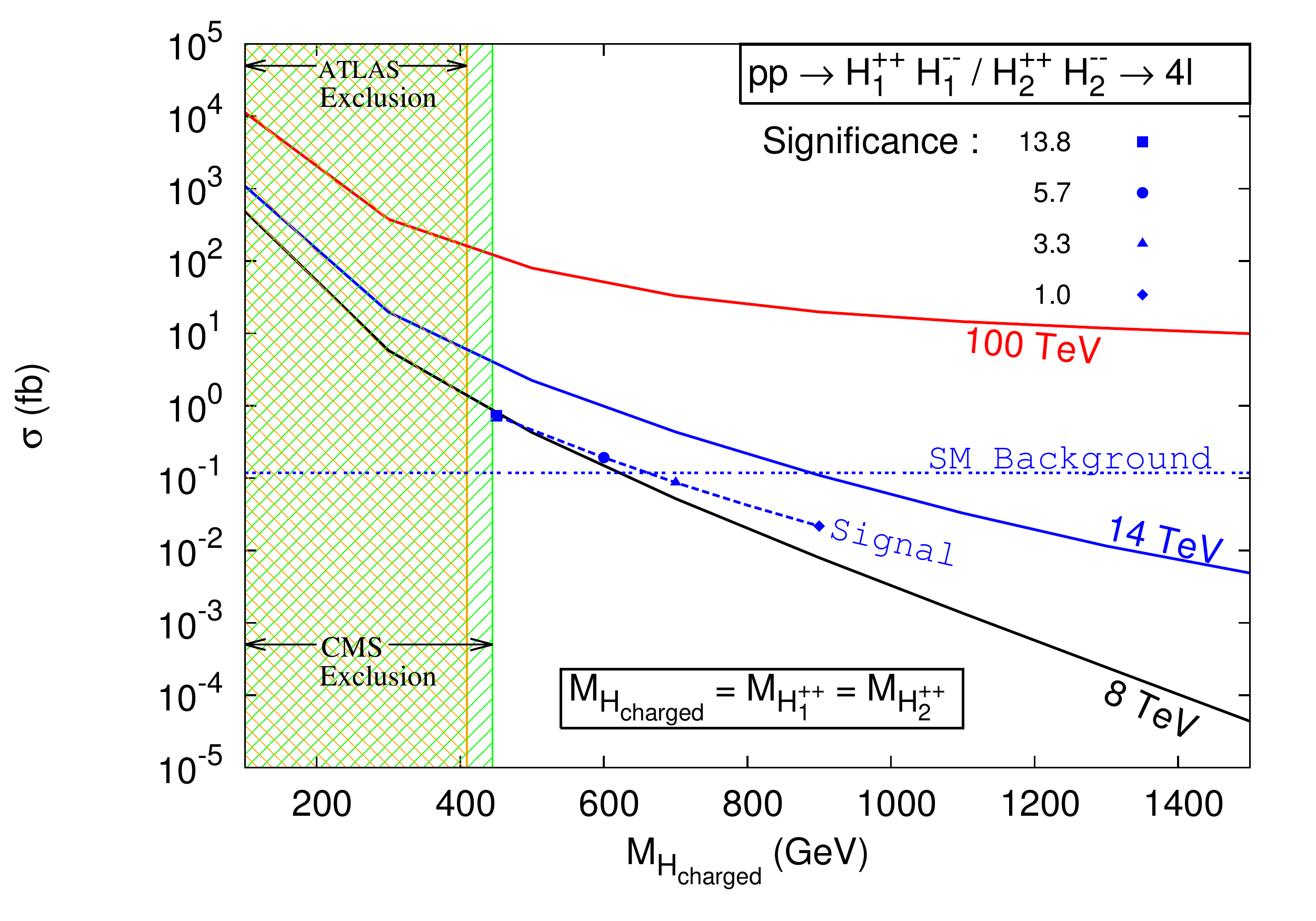,
angle= -90, scale=0.345}
\caption{\label{fig2}  
Cross-section $\sigma$ for the pair production of doubly charged scalars $H_i^{\pm\pm}$ decaying to four leptons for $\sqrt{s}=8$, $14$ and $100$ TeV at the LHC (solid lines) without the detailed selection cuts (see the text for details). The dotted-blue horizontal line is the estimation for the SM background for $4\ell$ final state with $\sqrt{s}=14$ TeV incorporating the detailed event selection criteria. Also 
the cross section for the  process $pp \to H_1^{++}H_1^{--}/H_2^{++}H_2^{--}\to 4 \ell$ 
with $\sqrt{s}=14$ TeV after implementing the same full selection cuts  is
depicted by the dashed-blue line for doubly charged scalar
mass range [450-900] GeV.  The four patches on the ``Signal" line denote the significance of the signal. It is assumed that $\delta\rho=4\rho_2$, what results in $M_{H_1^{\pm\pm}}=M_{H_2^{\pm\pm}}$, see Eqs. \eqref{B2MH00}-\eqref{B2MH2pp}. 
}
\end{figure}

Fig.~\ref{fig2} gives estimation of cross sections for the pair production of doubly charged Higgs bosons at LHC for the  center of mass energies at present (8 TeV), forthcoming run ($13 -14$ TeV) and in the further perspective (100 TeV). For the simplicity, we assume that masses of $H_1^{\pm\pm}$ and $H_2^{\pm\pm}$ are degenerate i.e. $\delta\rho=4\rho_2$. Limits on the doubly charged Higgs boson $H_1^{\pm\pm}$ masses from recent LHC data have been taken into account, and exclusion limits are explicitly given in Fig.~\ref{fig2}.

Here, the cross section for the process 
$pp \to (H_1^{\pm\pm} H_1^{\mp \mp} \oplus H_2^{\pm\pm} H_2^{\mp \mp}) \to 4\ell$
for different centre of mass energies $\sqrt{s}=8$, $14$ and $100$ TeV at the LHC with only basic cut (lepton $p_T>10$ GeV) is shown in Fig.~\ref{fig2} for different set of doubly charged scalar masses. We have also implemented the detailed event selection criteria \footnote{These event selection criteria are implemented in PYTHIA. They include lepton transverse momentum ($p_T$), pseudorapidity cuts, smearing of leptons, lepton-lepton (photon, jet) separation, hadronic activity, missing $p_T$ and Z-veto.}, as given in \cite{Bambhaniya:2013yca, Bambhaniya:2013wza} to compute the SM background for this final state coming from $t\bar{t} (Z/\gamma^*), (Z/\gamma^*) (Z/\gamma^*)$. Using the same set of cuts the signal event cross section is estimated for doubly charged scalar masses [450-900] GeV. 
We have pointed out four benchmark points for the following masses of doubly charged scalars: $450$, $600$, $700$, and $900\,\mathrm{GeV}$. The corresponding significances \footnote{Significance of the signal  is conveniently measured by the ratio $S/\sqrt{S+B}$, where $S$ and $B$ are the total number of signal and background events respectively.} of the signal for integrated luminosity $300$ fb$^{-1}$ are $13.8$, $5.7$, $3.3$, $1.0$ respectively. 
Thus at the LHC with $\sqrt{s}=14\,\mathrm{TeV}$ centre of mass energy and $300$ fb$^{-1}$ luminosity up to 600 GeV mass of the doubly charged scalars can be probed with significance~$\geq 5$. Finally, it is worthwhile to note that due to kinematics 
for $\sqrt{s}=100\,\mathrm{TeV}$ discussed 
cross-section does not change significantly when mass of doubly charged scalars increases. 

In summary, though Beyond Standard Model scenarios (to which MLRSM belongs) usually include mass spectrum at large energy scales,
relatively light scalar masses (below TeV level) might be allowed. In this report we have analysed correlations between the scalar masses and the scalar potential parameters within MLRSM framework. We have noted that it is still possible to find some of the scalar fields as light as few hundred GeV even when the new scale is around 
$\sim 10$ TeV, and these spectra are in good agreement with sharp conditions forced by experimental data. 
Though there are many free parameters in the Higgs sector, present experimental limits on masses of doubly 
charged Higgs bosons put already first partial constraints on some of these parameters. 
Moreover, our analysis shows that the lowest limit on doubly charged scalar mass $M_{H_{1}^{\pm\pm}}$ 
can be obtained. The other doubly charged scalar $H_2^{\pm\pm}$ is theoretically not constrained from below in the discussed model and it can be tuned to any desirable value, in agreement with experimental data. The lowest limit on $M_{H_1^{\pm\pm}}$ is important, as we know that production of $H_1^{\pm\pm}$ is much bigger than that of $H_2^{\pm\pm}$ \cite{Bambhaniya:2013wza}. 
%
%
%
Cross sections for the pair productions of doubly charged scalars followed by their leptonic decays have been discussed 
toghether with estimation of the SM background at 14 TeV centre of mass energy.

Using an example of the MLRSM model we have shown that we are slowly approaching era at the LHC physics where details of the elusive Higgs sector of the tested theory can be analysed.

\vspace{1cm}
\begin{acknowledgments}
This work was supported by Department of Science \& Technology, Government of India under the Grant Agreement No. IFA12-PH-34 (INSPIRE Faculty Award), the Research Executive Agency (REA) of the European Union
 under the Grant Agreement No. PITN-GA-2010-264564 (LHCPhenoNet) and Polish National Science Centre (NCN) under the Grant Agreement No.  DEC-2013/11/B/ST2/04023 and under postdoctoral grant No. DEC-2012/04/S/ST2/00003. MK is supported by the Forszt project co-financed by the EU from the
European Social Fund.
\end{acknowledgments}
 

\begin{thebibliography}{10}

\bibitem{Ellis:2013jnq}
J.~Ellis, 
\href{http://xxx.lanl.gov/abs/1312.5672}{arXiv:1312.5672}.

\bibitem{Aad:2012tfa}
G.~Aad et~al. (ATLAS Collaboration), 
{\em Phys.Lett.} {\bf B716} (2012) 1--29,
  [\href{http://xxx.lanl.gov/abs/1207.7214}{arXiv:1207.7214}].

\bibitem{Chatrchyan:2012ufa}
S.~Chatrchyan et~al. (CMS Collaboration), 
  {\em Phys.Lett.} {\bf B716} (2012) 30--61,
  [\href{http://xxx.lanl.gov/abs/1207.7235}{arXiv:1207.7235}].

\bibitem{Quigg:2009vq}
C.~Quigg, 
{\em Ann.Rev.Nucl.Part.Sci.} {\bf 59} (2009) 505--555,
[\href{http://xxx.lanl.gov/abs/0905.3187}{arXiv:0905.3187}].

\bibitem{Buttazzo:2013uya}
D.~Buttazzo, G.~Degrassi, P.~P. Giardino, G.~F. Giudice, F.~Sala, et~al., 
  {\em JHEP} {\bf
  1312} (2013) 089, [\href{http://xxx.lanl.gov/abs/1307.3536}{
  arXiv:1307.3536}].

\bibitem{Branchina:2013jra}
  V.~Branchina and E.~Messina,
  {\em Phys.\ Rev.\ Lett.} {\bf 111} (2013) 241801 
  [\href{http://xxx.lanl.gov/abs/1307.5193}{arXiv:1307.5193}].

\bibitem{Branchina:2014usa}
  V.~Branchina, E.~Messina and A.~Platania,
  {\em JHEP} {\bf 1409} (2014) 182 
  [\href{http://xxx.lanl.gov/abs/1407.4112}{arXiv:1407.4112}].

\bibitem{Branchina:2014rva}
  V.~Branchina, E.~Messina and M.~Sher,
  {\em Phys.\ Rev.} {\bf D91} (2015) 1,  013003 
  [\href{http://xxx.lanl.gov/abs/1408.5302}{arXiv:1408.5302}].

\bibitem{Jegerlehner:2013cta}
F.~Jegerlehner, 
  {\em Acta Phys.\ Polon.} {\bf B45} (2014) 1167, 
  [\href{http://xxx.lanl.gov/abs/1304.7813}{arXiv:1304.7813}].

\bibitem{Mohapatra:1974gc}
R.~Mohapatra and J.~C. Pati, 
{\em
  Phys.Rev.} {\bf D11} (1975) 2558.

\bibitem{Senjanovic:1975rk}
G.~Senjanovic and R.~N. Mohapatra, 
{\em Phys.Rev.} {\bf D12} (1975) 1502.

\bibitem{Duka:1999uc}
P.~Duka, J.~Gluza, and M.~Zralek, 
{\em
  Annals Phys.} {\bf 280} (2000) 336--408,
  [\href{http://xxx.lanl.gov/abs/hep-ph/9910279}{hep-ph/9910279}].

\bibitem{Mohapatra:1986uf}
R.~Mohapatra, {\it {Unification and Supersymmetry. The Frontiers of Quark-Lepton Physics}} (Springer, New York, 1986).

\bibitem{CMS:2012kua}
CMS Collaboration, 
  Report No. \rm{CMS-PAS-HIG-12-005}.

\bibitem{ATLAS:2012hi}
G.~Aad et~al. (ATLAS Collaboration), 
{\em Eur.Phys.J.} {\bf C72} (2012) 2244,
  [\href{http://xxx.lanl.gov/abs/1210.5070}{arXiv:1210.5070}].
  
\bibitem{Bambhaniya:2013wza}
G.~Bambhaniya, J.~Chakrabortty, J.~Gluza, M.~Kordiaczynska, and R.~Szafron,
{\em
  JHEP} {\bf 1405} (2014) 033, [\href{http://xxx.lanl.gov/abs/1311.4144}{
  arXiv:1311.4144}].
   
\bibitem{Melfo:2011nx}
  A.~Melfo, M.~Nemevsek, F.~Nesti, G.~Senjanovic and Y.~Zhang,
 {\em  Phys.Rev.} {\bf D85} (2012) 055018,
  [\href{http://xxx.lanl.gov/abs/1108.4416}{arXiv:1108.4416}].  

\bibitem{kang:2014jia}
  Z.~Kang, J.~Li, T.~Li, Y.~Liu and G.~-Z.~Ning,
  [\href{http://xxx.lanl.gov/abs/1404.5207}{arXiv:1404.5207}].

\bibitem{Gunion:1989in}
J.~Gunion, J.~Grifols, A.~Mendez, B.~Kayser, and F.~I. Olness, 
{\em Phys.Rev.} {\bf D40} (1989)
  1546.

\bibitem{Gluza:1994ad}
J.~Gluza and M.~Zralek, 
{\em Phys.Rev.} {\bf
  D51} (1995) 4695--4706, [\href{http://xxx.lanl.gov/abs/hep-ph/9409225}{
  hep-ph/9409225}].

\bibitem{Beringer:1900zz}
J.~Beringer et~al. (Particle Data Group),  {\em Phys.Rev.} {\bf D86} 010001 (2012) and 2013 partial update for the 2014 edition. K.~A.~Olive et al.  (Particle Data Group Collaboration),
{\em Chin.\ Phys.} {\bf C 38} (2014) 090001.

\bibitem{Ecker:1983uh}
G.~Ecker, W.~Grimus, and H.~Neufeld, 
{\em Phys.Lett.} {\bf
  B127} (1983) 365.

\bibitem{Mohapatra:1983ae}
R.~N. Mohapatra, G.~Senjanovic, and M.~D. Tran, 
  {\em Phys.Rev.} {\bf D28} (1983) 546.

\bibitem{Pospelov:1996fq}
M.~Pospelov, 
{\em Phys.Rev.} {\bf D56} (1997) 259--264,
  [\href{http://xxx.lanl.gov/abs/hep-ph/9611422}{hep-ph/9611422}].
  
  \bibitem{Zhang:2007da}
  Y.~Zhang, H.~An, X.~Ji and R.~N.~Mohapatra,
  {\em Nucl.Phys.} {\bf B802} (2008) 247,
  [\href{http://xxx.lanl.gov/abs/0712.4218}{arXiv:0712.4218}].

\bibitem{Maiezza:2010ic}
A.~Maiezza, M.~Nemevsek, F.~Nesti, and G.~Senjanovic, 
{\em Phys.Rev.} {\bf D82} (2010) 055022,
  [\href{http://xxx.lanl.gov/abs/1005.5160}{arXiv:1005.5160}].

\bibitem{Chakrabortty:2012pp}
J.~Chakrabortty, J.~Gluza, R.~Sevillano, and R.~Szafron, 
{\em JHEP} {\bf 1207}
  (2012) 038, [\href{http://xxx.lanl.gov/abs/1204.0736}{
  arXiv:1204.0736}].

\bibitem{Bertolini:2014sua}
  S.~Bertolini, A.~Maiezza and F.~Nesti,
  {\em Phys.Rev.} {\bf D89} (2014) 095028,
  [\href{http://xxx.lanl.gov/abs/1403.7112}{arXiv:1403.7112}].
  
\bibitem{Guadagnoli:2010sd}
D.~Guadagnoli and R.~N. Mohapatra, 
  {\em Phys.Lett.} {\bf B694} (2011)
  386--392, [\href{http://xxx.lanl.gov/abs/1008.1074}{arXiv:1008.1074}].

\bibitem{CMS:2012uaa}
CMS Collaboration, 
 Report No. \rm{CMS-PAS-EXO-12-017}.

\bibitem{salo:1558322}
C.~Vuosalo (CMS Collaboration), 
Report No. \rm{CMS-CR-2013-155}; 
  C.~Vuosalo (CMS Collaboration),
  PoS DIS 2013 (2013) 122
  \href{http://xxx.lanl.gov/abs/1307.2168}{[arXiv:1307.2168]}.

\bibitem{Deppisch:2014qpa}
  F.~F. Deppisch, T.~E. Gonzalo, S.~Patra, N.~Sahu, and U.~Sarkar, 
  Phys.\ Rev. {\bf D90} (2014) 5,  053014
  \href{http://xxx.lanl.gov/abs/1407.5384}{[arXiv:1407.5384]}.

\bibitem{Heikinheimo:2014tba}
  M.~Heikinheimo, M.~Raidal and C.~Spethmann,
  \href{http://xxx.lanl.gov/abs/1407.6908}{arXiv:1407.6908}.  

\bibitem{Datta:1999nc}
A.~Datta and A.~Raychaudhuri, 
{\em
  Phys.Rev.} {\bf D62} (2000) 055002,
  [\href{http://xxx.lanl.gov/abs/hep-ph/9905421}{hep-ph/9905421}].

\bibitem{Czakon:1999ue}
M.~Czakon, M.~Zralek, and J.~Gluza, 
{\em Nucl.Phys.} {\bf B573} (2000) 57--74,
  [\href{http://xxx.lanl.gov/abs/hep-ph/9906356}{hep-ph/9906356}].

\bibitem{Czakon:2002wm}
M.~Czakon, J.~Gluza, and J.~Hejczyk, 
{\em Nucl.Phys.} {\bf B642} (2002) 157--172,
  [\href{http://xxx.lanl.gov/abs/hep-ph/0205303}{hep-ph/0205303}].

\bibitem{Bambhaniya:2013yca}
G.~Bambhaniya, J.~Chakrabortty, S.~Goswami, and P.~Konar, 
{\em Phys.Rev.} {\bf D88} (2013), no.~7 075006,
  [\href{http://xxx.lanl.gov/abs/1305.2795}{arXiv:1305.2795}].
  
 
\end{thebibliography}

\end{document}